\documentclass[runningheads]{llncs}
\usepackage[T1]{fontenc}
\usepackage{graphicx}
\usepackage{subcaption}
\usepackage{graphicx} 
\usepackage{siunitx}
\usepackage{dsfont}
\usepackage[table,dvipsnames]{xcolor}
\usepackage{siunitx}

\begin{document}
\title{Fine-tuned Normalizing Flows for ALICE Zero Degree Calorimeter Fast Simulation}
\titlerunning{Fine-tuned Normalizing Flows for ZDC Fast Simulation}
%
\author{Emilia Majerz\inst{1}\orcidID{0009-0005-2034-0410}\thanks{Corresponding author: majerz@agh.edu.pl} \and
Jacek Otwinowski\inst{2}\orcidID{0000-0002-5471-6595} \and
Witold Dzwinel\inst{1}\orcidID{0000-0001-8321-5928}  \and
Jacek Kitowski\inst{1}\orcidID{0000-0003-3902-8310} \\}
\authorrunning{E. Majerz et al.}
%
\institute{
AGH University of Krakow, Poland
\and
The Henryk Niewodniczański Institute of Nuclear Physics, Polish Academy of Sciences}
%
\maketitle              
\begin{abstract}
Simulating the ALICE Zero Degree Calorimeter (ZDC) neutron detector responses at the LHC is computationally expensive, requiring complex Monte Carlo chains. We develop a generative surrogate, focusing on Normalizing Flows (NFs). Through transfer learning, we pre-train on the full imbalanced dataset and fine-tune specialized models for different particle types ($\gamma$, $n$, $\Lambda$, $K_S^0$, $\Sigma^+$) using two gradual-unfreezing schemes. As standard ZDC metrics like Wasserstein distance overlook conditional structure, we introduce refined metrics: conditional weighted MAE, dispersion ratio, and Jaccard co-activation error, that better capture physics-relevant input-output dependencies and response variability. Our ensemble of fine-tuned models achieves a Wasserstein distance of $1.61 \pm 0.02$, outperforming baselines across all metrics. This work provides a generalizable NF-based framework for LHC detector simulation, combining NFs, conditional fine-tuning, and physics-motivated evaluation.




\end{abstract}
\section{Introduction}
\label{sec:intro}

The simulation of particle detectors at the LHC traditionally employs computationally demanding Monte Carlo simulation paradigms. This complexity restricts the volume of simulated collision data available for interpreting and calibrating real-world experimental data, driving researchers to pursue alternative methodologies. With their recent advances, fast simulations employing Generative Neural Networks offer a promising alternative as a baseline for numerical models' surrogates. 

\begin{figure}[ht]
\centering
\includegraphics[width=0.75\linewidth]{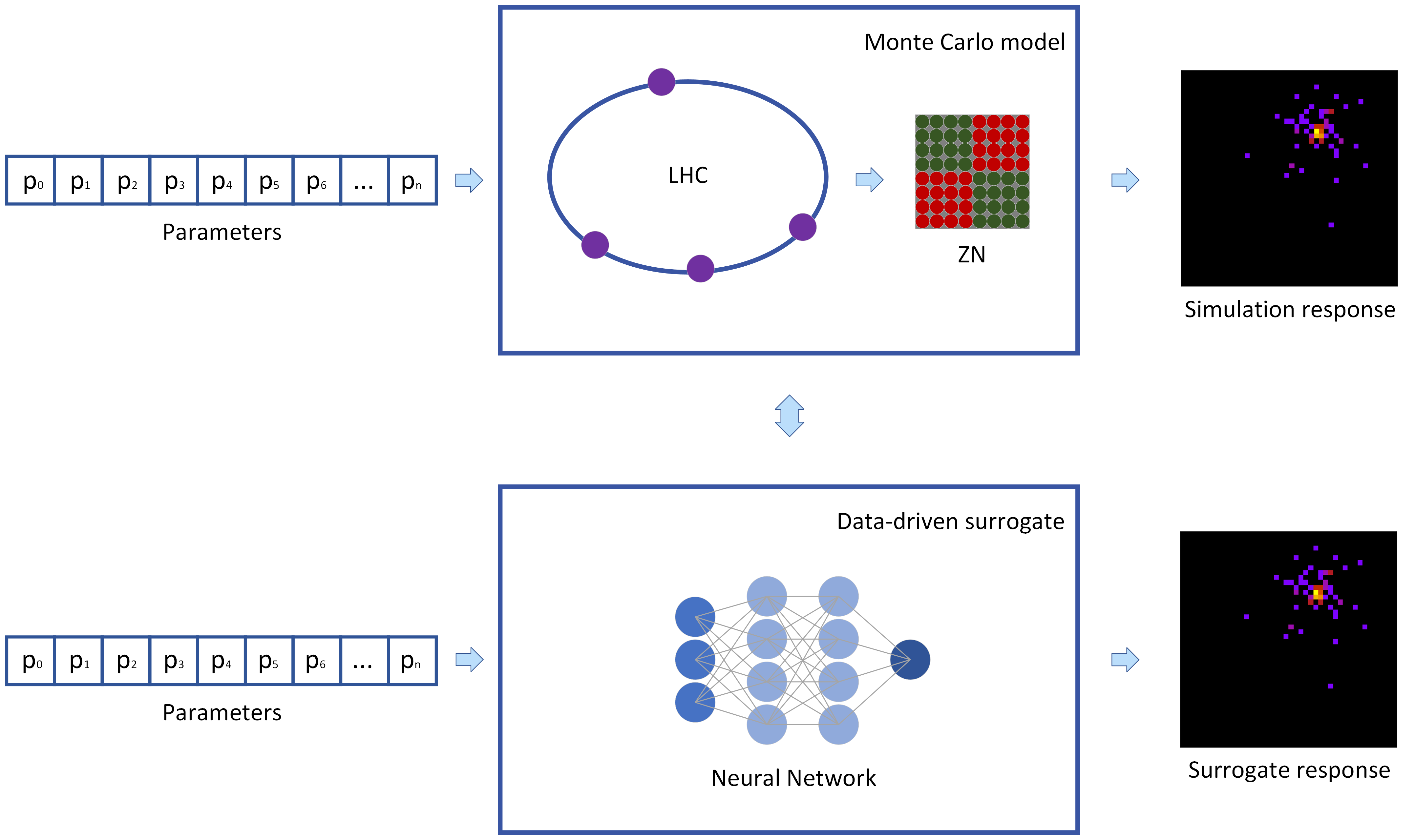}
\caption{The relation between numerical simulation (Monte Carlo model) and neural network-based surrogate approach utilised in fast simulations.}
\label{fig:experiment-overview}
\end{figure}

The Zero Degree Calorimeter (ZDC) of the ALICE experiment~\cite{Gallio:381433} is a particularly relevant target for fast simulation. Located \SI{112.5}{\metre} from the Interaction Point 2 (IP2), it requires many steps of the Monte Carlo method to obtain the simulated response. The ZDC consists of two detectors on each side of IP2: the proton calorimeter (ZP) and the neutron calorimeter (ZN). The ZN contains a quartz fibre matrix of size 44 $\times$ 44 embedded in a high-density tungsten absorber, with dimensions of \SI{7.04} \times \SI{7.04} \times \SI{100}{\centi\metre\cubed}. When a particle enters the ZN, it initiates hadronic showers in the absorber, and the resulting charged particles produce Cherenkov light in the fibres. The readout is organised into four towers in a 2 $\times$ 2 layout (depicted as red and green squares in Fig.~\ref{fig:experiment-overview}, presenting the experiment overview). The output of every other fibre is routed to a shared photomultiplier (PM), with four additional PMs collecting signals from each of the respective towers~\cite{Gallio:381433}. This structure makes the ZN a natural test case for image-based surrogate modelling and for validating physically meaningful predictions.


\vspace{-15pt}

\vspace{-15pt}

Physics-Based Deep Learning (PBDL), also referred to as Scientific Machine Learning, combines physical modelling, numerical simulation, and modern deep learning. Here, we propose a supervised data-driven approach to build a fast surrogate for developing a fast surrogate of complex numerical models.

Fast calorimeter simulation using generative models has been widely studied. Early approaches used Generative Adversarial Networks~\cite{krause2024calochallenge2022communitychallenge} and autoencoders~\cite{krause2024calochallenge2022communitychallenge}. More recent work has explored Normalizing Flows (NFs)~\cite{Krause_2023,krause2024calochallenge2022communitychallenge,10.21468/SciPostPhys.17.2.045}, diffusion models~\cite{krause2024calochallenge2022communitychallenge}, and flow matching~\cite{krause2024calochallenge2022communitychallenge,WOJNAR2026109936}. Several studies have also addressed ZN simulation specifically~\cite{9311504,Dubinski2023MachineLM,kita2024generativediffusionmodelsfast,zdcfastsim,10.1007/978-3-031-30105-6_22,WOJNAR2026109936,Majerz:2025ykn}, but only few have investigated NFs for this detector~\cite{zdcfastsim,Majerz:2025ykn}.

NFs are a suitable choice for this task because they provide exact likelihood estimation, invertibility, and strong expressive power. These properties are attractive in a physics setting, where predictive accuracy should be accompanied by interpretability and consistency with the underlying detector response. NFs are also well suited to tasks with substantial conditional variability, where the same input can correspond to multiple valid outputs, as in the ZN dataset.

The dataset is highly imbalanced: common particle types account for most responses, whereas rarer species are underrepresented. While this distribution reflects realistic physical conditions, it poses challenges for training predictive models, which may become biased toward dominant particle types. To address this, we pre-train a base model on the full multi-particle dataset and then fine-tune dedicated submodels for selected particle types using gradual unfreezing. During inference, the model corresponding to the input particle type is used, which improves performance across both frequent and rare categories. Although fine-tuning methods for NFs have been proposed~\cite{gambardella2019transflow,siahkoohi2020fasteruncertaintyquantificationinverse,Malnick_2024_WACV,10.21468/SciPostPhys.17.2.045}, to the best of our knowledge they have not been systematically evaluated for MAF-based simulators in a high-energy physics setting using physics-motivated metrics. In addition, we assess model behaviour with interpretability methods to check whether the learned dependencies are physically plausible.


Our work concerns the application of NFs in the simulation of the ZN responses, with a particular focus on transfer learning and fine-tuning. We extend the research on NFs presented in~\cite{zdcfastsim}, and also make the following contributions:
\begin{enumerate}
    \item We improve the MAF-based NF surrogate through fine-tuning and an ensemble approach.
    \item We propose and compare two schemes for MAF-based NFs fine-tuning, utilising the gradual unfreezing technique.
    \item We propose and compare four methods for noise removal after NFs training, observing significant differences in model performance.
    \item After carrying out a critical analysis of the metrics reported in the literature for ZDC surrogate evaluation, we identify their weaknesses, and propose refined variants.
    \item We use SHAP~\cite{NIPS2017_8a20a862} to assess whether the model relies on physically meaningful input variables, and show that the model’s predictions are primarily influenced by energy and momentum components, which is consistent with known detector response patterns.
\end{enumerate}

The rest of the paper is organised as follows. Section~\ref{sec:dataset-challenges} describes the dataset, Section~\ref{sec:methodology} presents the methodology, and Section~\ref{sec:results} reports the results and comparison with prior work. We conclude with an interpretability analysis of the ensemble.



\section{Dataset}
\label{sec:dataset-challenges}

The dataset used in this study contains N=$306{,}780$ samples generated by the GEANT4~\cite{AGOSTINELLI2003250} software. Each sample is a pair consisting of a vector with features of particles produced by primary proton-proton collisions (called primary particles) -- such as energy, three-momentum, position in three dimensions, mass, and charge -- and a $44 \times 44$ matrix representing the ZN response. Each matrix element represents the number of photons detected in the corresponding fibre. The detector response is therefore treated as a single-channel image, where each pixel value corresponds to a photon count. To suppress low-signal random responses, we retain only matrices with at least 10 photons; after this filtering, the median photon count is 234.


An important property of the dataset is its strong class imbalance. The $\gamma$ particles account for almost 61\% of the data, $n$ for over 23\%, $p$ for over 5\%, $\Lambda$ for over 3\%, and the remaining 17 particle types together for around 7.5\%. This imbalance may bias the model toward the dominant particle types and degrade performance on underrepresented ones. Possible remedies include rebalancing the dataset, applying standard techniques for imbalanced data~\cite{kubat1997addressing}, or training dedicated models for poorly represented particle types. Here, we investigate the last strategy.


Among the particles used in our experiments, the fractions of primary particles that hit the detector directly and produce a response of at least 10 photons are 45\% for $\gamma$, 78\% for $n$, 46\% for $\Lambda$, 21\% for $K_S^0$, and 0\% for $\Sigma^+$. When the fractions are lower, we should expect multiple showers and scattered responses induced by secondaries.

\begin{figure}[ht]
\centering
\includegraphics[width=0.65\linewidth,trim={0 5cm 0 0},clip]{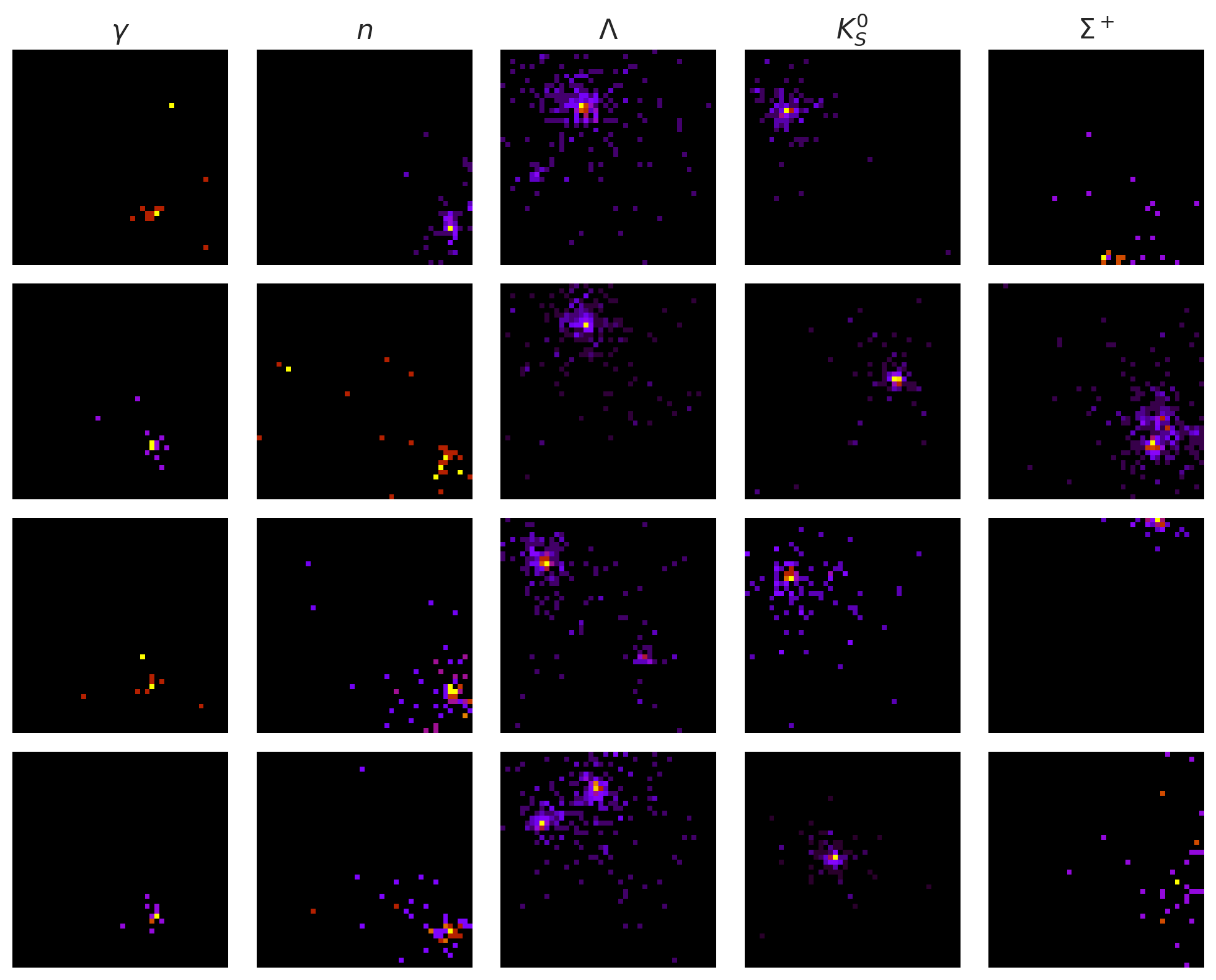}
\caption{The illustration of the diversity of the detector responses. Each column represents three GEANT4 simulations of the same input particle.}
\label{fig:diverse-responses}
\end{figure}

The diversity of detector responses is an additional challenge for any data-driven surrogate model. The same vector of input parameters can produce substantially different ZN responses. This happens because particles traverse a long distance between IP2 and the detector, during which secondary particles may be produced and primary trajectories may deviate from their original paths. Different particle types also exhibit different levels of response diversity: some produce fairly consistent showers, while others lead to much more variable shower positions and shapes. An illustrative example is shown in Fig.~\ref{fig:diverse-responses}. To quantify this effect, we use the variance-based diversity measure introduced in~\cite{10.1007/978-3-031-30105-6_22}:
\begin{equation}\label{eq-div}
    f_{\mathrm{div}}(\mathbf{c}) = \sum_{i,j} \sqrt{\frac{1}{|X_c|}\sum_{t \in X_c}\left(x^t_{ij} - \mu_{ij}\right)^2},
\end{equation}
where $x^t_{ij}$ is the pixel value at coordinates $i,j$ for the detector response $t$, $\mu_{ij}$ is the mean pixel value at $i,j$ for the unique input vector $\mathbf{c}$, and $|X_c|$ is the number of detector responses associated with input $c$. This quantity is positively correlated with particle mass and energy.

\vspace{-10pt}



\section{Methodology}
\label{sec:methodology}

We use a three-stage approach:
\vspace{-5pt}
\begin{enumerate}
    \item Pre-train an NF model on the full, imbalanced dataset to capture general detector-response patterns.
    \item Fine-tune the pre-trained model separately for selected particle types using the corresponding subsets.
    \item Combine the fine-tuned models into an ensemble and verify its performance.
\end{enumerate}


We split the dataset into training, validation, and test sets in a 70:10:20 ratio. Models with the best validation performance were selected for testing. During fine-tuning, we used only particle-specific samples from the corresponding splits to avoid data leakage. Each model was evaluated five times, generating five samples per input vector on the test set.




\subsection{Two-stage modelling}\label{sec:two-stage}

In our initial experiments, we trained NFs directly on the primary particle features. To reduce the dynamic range of the detector-response images, we applied a log transform to the pixel values. Although the generated showers were visually plausible, the predicted photon counts did not match the data well. We therefore adopted a two-stage approach: the NF generates a normalized response shape, which is then rescaled by an auxiliary model predicting the expected total photon count. We also provided the expected photon count as an additional input feature during NF training, following~\cite{Krause_2023}.

We compared several models for predicting the photon sum, including NFs, SVMs, GBRs, and BNNs. The BNN performed best and was used in all experiments (MAE: 53.20, RMSE: 111.47). Although the absolute errors appear large, the median percentage error of the best BNN was 12\%, which seems acceptable given the stochastic nature of the target.



\begin{figure}[ht]
\centering
\includegraphics[width=\linewidth]{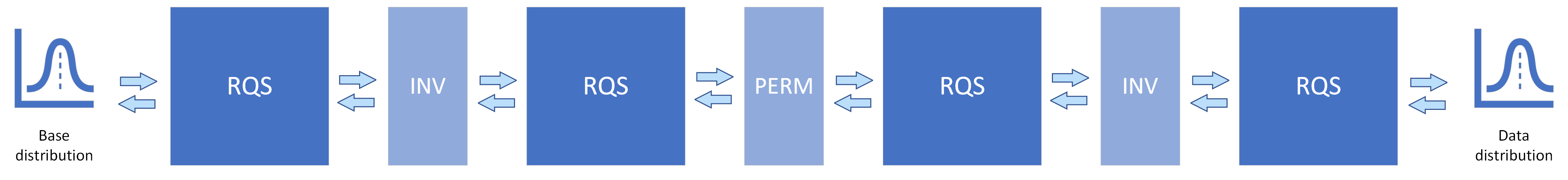}
\caption{Block diagram of the NF model used. RQS layers are interleaved with permutation (PERM) and inversion (INV - order reversal) layers.}
\label{fig:nf-block-diagram}
\end{figure}

For the NF model, we follow the configuration of~\cite{zdcfastsim}, itself based on~\cite{Krause_2023}. The model uses the MAF architecture with rational quadratic spline (RQS) transformations~\cite{d1220ead849147929d687c8a42669907}, whose parameters are produced by MADE blocks~\cite{germain2015mademaskedautoencoderdistribution}. The NF maps the data distribution to a simple base distribution through a sequence of invertible transformations (Fig.~\ref{fig:nf-block-diagram}). Sampling proceeds in the reverse direction: a point is drawn from the base distribution, the MADE blocks compute the spline parameters, and the transformations are applied successively. Permutation layers are inserted between spline blocks to increase flexibility. Compared with~\cite{zdcfastsim}, we improved the baseline by replacing the conditional diagonal Normal base with a standard Normal and using the pixel-value recalculation post-processing scheme.


\subsection{Performance metrics}\label{subsec:performance_metrics}


Because multiple detector responses may be valid for the same input, direct pixel-wise comparison is not meaningful. Following prior work on the ZN detector~\cite{9311504,Dubinski2023MachineLM,kita2024generativediffusionmodelsfast,zdcfastsim,10.1007/978-3-031-30105-6_22}, we use the mean 1-Wasserstein distance between the distributions of photons collected by each PM. We refer to these groups as channels and report the average over $m=5$ channels:
\vspace{-5pt}
\begin{equation}\label{wasserstein}
    W_1(w, \hat{w}) = \frac{1}{m}\sum_{i=1}^{m}\int_0^1 \left|F^{-1}_{w_i}(z)-F^{-1}_{\hat{w}_i}(z)\right|\,dz.
\end{equation}
Here, $F^{-1}_{w_i}$ is the inverse cumulative distribution function of channel distribution $w_i$, and $\hat{w}_i$ denotes the predicted distribution.



\begin{figure}[ht]
\centering
\includegraphics[width=0.75\linewidth]{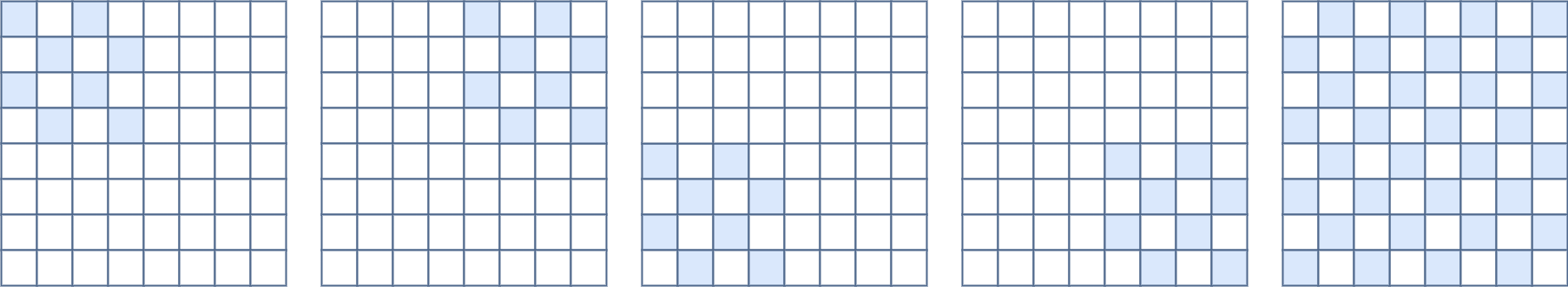}
\caption{A schematic overview of checkerboard masks applied to detector responses to obtain channel values. Each mask is the same size as the response.}
\label{fig:channels-checkerboards}
\end{figure}

We compute channel values by applying five checkerboard masks to each response image and summing the values within predefined fibre groups. A schematic overview of the masks is shown in Fig.~\ref{fig:channels-checkerboards}.


\vspace{-10pt}
\vspace{-10pt}



Although the Wasserstein distance captures global distributional similarity, it does not assess whether the response matches the input condition. Thus, we compute a conditional mean absolute error over unique inputs, as in~\cite{Majerz:2025ykn}:
\vspace{-5pt}
\begin{equation}\label{mae-cond}
    \mathrm{MAE}_{cond}(w,\hat{w}) = \frac{1}{\sum_k p_k}\sum_{k=1}^{c} p_k \cdot \frac{1}{m}\sum_{i=1}^{m}\left|\overline{w_i^k}-\overline{\hat{w}_i^k}\right|.
\end{equation}
We use logarithmic weights, $p_k=\log(1+n_k)$, to reduce the influence of highly repeated conditions while still favouring well-sampled regions of phase space.


To quantify channel variability, we use a per-channel median absolute deviation (MAD) with shrinkage toward a particle-level prior. MAD is robust to outliers, which is useful for detector responses that may contain rare, extreme showers, and shrinkage further stabilises it in low-sample cases. We compute the prior $MAD_{particle}$ as a weighted median MAD value computed for reference inputs with at least 5 responses available. For each unique input and channel:
\begin{equation}
    \mathrm{MAD}_{u,c} = \mathrm{median}_{i}\left(\left|R^i_{u,c}-\mathrm{median}_{j}(R^j_{u,c})\right|\right),
\end{equation}
\vspace{-15pt}
\begin{equation}
    \widetilde{\mathrm{MAD}}_{u,c} = \alpha_u \,\mathrm{MAD}_{u,c} + (1-\alpha_u)\,\mathrm{MAD}_{particle,c},
    \qquad
    \alpha_u = \frac{N_u}{N_u+\kappa}.
\end{equation}
Here, $\kappa$ is the 10th percentile of the number of responses per particle type. We compare reference and generated variability using the dispersion ratio:
\begin{equation}\label{delta}
    \delta_{u,c} = \frac{\left|\widetilde{\mathrm{MAD}}_{u,c}^{gen}-\widetilde{\mathrm{MAD}}_{u,c}^{ref}\right|}{\widetilde{\mathrm{MAD}}_{u,c}^{gen}+\widetilde{\mathrm{MAD}}_{u,c}^{ref}+\tau},
\end{equation}
which provides a bounded symmetric score, treating overdispersion and underdispersion equally. We compute this separately for each channel and report the mean over channels, followed by a weighted mean over unique inputs. Cases with fewer than two samples are excluded.

We also measure spatial co-activation between neighbouring channels using a Jaccard-based error. For each event, we define a binary hit indicator and compute the Jaccard index for each neighbouring pair $(c_1,c_2)$:
\begin{equation}
    J_{c_1,c_2} = \frac{\sum_i \mathds{1}(B^i_{u,c_1}=1 \wedge B^i_{u,c_2}=1)}{\sum_i \mathds{1}(B^i_{u,c_1}=1 \vee B^i_{u,c_2}=1)}, \qquad B^i_{u,c}=\mathds{1}(R^i_{u,c}>0).
\end{equation}
If both channels are always zero, we set $J_{c_1,c_2}=1$. The final score for a unique input is
\begin{equation}\label{JaccErr}
    \mathrm{JaccErr}_u = \frac{1}{P}\sum_{\text{pairs}}\left|J_u^{gen}-J_u^{ref}\right|,
\end{equation}
where $P$ is the number of neighbouring channel pairs. We report the weighted mean over unique inputs.

\subsection{Fine-tuning Normalizing Flows}\label{sec:fine-tuning}

After training on the full dataset, we fine-tune separate models for selected particle types. We focus on particle types for which the baseline Wasserstein score is relatively poor and exclude very rare classes from the fine-tuning study. The selected particles are $\gamma$, $n$, $\Lambda$, $K_S^0$, and $\Sigma^+$. Although fine-tuning all particle types might yield further gains, our goal here is to validate the methodology rather than to optimise every class exhaustively.


Standard neural-network fine-tuning usually freezes early layers and retrains later ones. For NFs, we consider two gradual-unfreezing directions: from the base-distribution side toward the data side, and in the opposite direction. We evaluate both because the flow structure makes either choice plausible. In each step, we unfreeze one MADE block at a time.



\subsection{Post-processing}\label{subsec:postprocessing}

Discrete-valued data, such as detector responses, are often modelled by adding small uniform noise so that each integer value corresponds to a small interval. We use the same idea here. In the two-stage setup, the NF first generates a normalized response shape, which is then rescaled by the predicted total photon count (including the noise contribution). After this rescaling, we remove the added noise using one of four schemes:
\begin{enumerate}
    \item\label{floorop1} Apply the floor function.
    \item\label{subop1} Subtract the expected preprocessing noise and round to the nearest integer.
    \item\label{floorop2} Apply the floor function and then renormalize and rescale the response to preserve the total photon count. Finish with rounding.
    \item\label{subop2} Subtract the expected preprocessing noise, round, and then renormalize and rescale. Finish with rounding.
\end{enumerate}
The renormalization steps are used because direct floor or round operations may force some pixels to zero and distort the total photon count. After renormalization, we rescale the responses, and round the values to integers again.

\subsection{Ensemble model}

The fine-tuned models can be combined with the baseline model in an ensemble. For each input, we use the dedicated fine-tuned model if one exists; otherwise, we use the baseline model trained on the full dataset. The ensemble therefore assigns specialised models to particle types that benefit from fine-tuning while retaining a shared model for the rest. In principle, the submodels may also differ in architecture, but in this work we use a uniform MAF-based setup.


\section{Results}
\label{sec:results}

We report means and standard deviations over five runs of generating responses on the test set, which covers 20\% of the dataset.

\subsection{Post-processing approaches}

In Section~\ref{subsec:postprocessing}, we introduced four methods for noise removal. The chosen post-processing approach has a large effect on Wasserstein performance, and no single method is optimal across all particle types, as shown in Table~\ref{tab:results-postprocessing}. Method~\ref{floorop2} is best overall, but particle-specific optima differ. Using the default post-processing method, the Wasserstein score for the whole dataset is $1.73 \pm 0.02$. When a dedicated method is selected for each particle type, the score improves slightly to $1.70 \pm 0.02$.


\begin{table}[ht]
    \caption{Comparison of noise-removal methods on samples generated by the best-performing baseline model, quantified by the Wasserstein score.}
    \centering
    \begin{tabular}{ccccc}
        \hline
        & Method~\ref{floorop1} & Method~\ref{subop1} & Method~\ref{floorop2} & Method~\ref{subop2} \\
        \hline
        All & $ 2.60 \pm 0.03$ & $4.33 \pm 0.02$ & $\mathbf{1.73 \pm 0.02}$ & $2.46 \pm 0.02$\\
        Gamma ($\gamma$) & $3.67 \pm 0.02$ & $3.98 \pm 0.01$ & $\mathbf{1.80 \pm 0.02}$ & $1.92 \pm 0.02$ \\
        Neutron ($n$) & $\mathbf{3.47 \pm 0.07}$ & $9.55 \pm 0.03$ & $3.70 \pm 0.07$ & $5.96 \pm 0.06$\\
        Lambda ($\Lambda$) & $\mathbf{16.23 \pm 0.22}$ & $18.63 \pm 0.17$ & $17.39 \pm 0.24$ & $19.60 \pm 0.20$\\
        K-Short ($K_S^0$) & $12.52 \pm 0.12$ & $\mathbf{10.51 \pm 0.06}$ & $11.66 \pm 0.07$ & $11.44 \pm 0.11$\\
        Sigma+ ($\Sigma^+$) & $\mathbf{33.94 \pm 0.35}$ & $35.04 \pm 0.21$ & $40.70 \pm 0.16$ & $41.11 \pm 0.15$\\
        \hline
    \end{tabular}
    \label{tab:results-postprocessing}
\end{table}



\subsection{Transfer learning and fine-tuning}

We fine-tuned the baseline NF model (Baseline\_100) for $\gamma$, $n$ $\Lambda$, $K_s^0$, and $\Sigma+$ particles, separately. The model was first trained for 100 epochs and then fine-tuned for another 100 epochs using the two schemas introduced in Section~\ref{sec:fine-tuning}: FT\_100\_ND, where unfreezing starts from layers close to the Normal distribution, and FT\_100\_DN, where unfreezing starts from layers close to the data. We also compared the results with a baseline model trained for 200 epochs (Baseline\_200). In addition, we trained separate models from scratch for each particle type (Individual), with 100 and 200 epochs, but these performed substantially worse than the other approaches.

\subsubsection{Two-stage modelling: the NF model (second stage) analysis}\label{subsub:second-stage-analysis}

We first assess the NF models responsible for modelling shower position and shape independently of the model predicting the expected number of photons. To do this, we use the photon counts derived from the dataset as input.

The analysed particles differ substantially in sample size, in the average number of responses per unique input, and in their level of \textit{diversity} (Eq.~\ref{eq-div}), as provided in Table~\ref{tab:dataset-stats}. Here, particle \textit{diversity} is computed as a weighted mean of per-unique-input diversities normalised to $[0,1]$, with weights given by the number of responses available for each input.


\vspace{-15pt}
\begin{table}[ht]
    \caption{Basic statistics of the dataset with respect to the analysed particle types.}
    \centering
    \resizebox{\textwidth}{!}{%
    \begin{tabular}{cccccc}
        \hline
         & Gamma ($\gamma$) & Neutron ($n$) & Lambda ($\Lambda$) & K-Short ($K_S^0$) & Sigma+ ($\Sigma^+$) \\
         \hline
         No. of test samples & 37,329 & 14,310 & 1,861 & 1,305 & 247 \\
         No. of unique test inputs & 608 & 207 & 45 & 87 & 19 \\
         Diversity & 0.10 & 0.30 & 0.48 & 0.47 & 0.58 \\
         \hline
    \end{tabular}}
    \label{tab:dataset-stats}
\end{table}

Table~\ref{tab:results-orgps} shows the results of our experiments. None of the approaches is best in every case, but FT\_100\_ND stands out as the strongest overall. The $\Sigma^+$ case remains exploratory because only a small number of samples is available. With that caveat, FT\_100\_ND performs best in 3 out of 4 cases for the $\text{MAE}_{cond}$, $\delta$, and JaccErr metrics. These metrics are computed at the level of unique input vectors, preserving the physical input-output relationship. By contrast, the Wasserstein score is more global and can favour distributional similarity even when the conditional structure is less accurate. FT\_100\_DN performs better than the baseline in all metrics for the $\gamma$ case, which suggests that this fine-tuning direction can also be useful for some datasets.


\begin{table}[ht]
    \caption{Comparison of model performances for different particles and metrics, with input photon counts derived from the dataset. \textit{Baseline} denotes models trained on the full dataset and \textit{FT\_100} denotes fine-tuned Baseline\_100 models. The best results are in \textbf{bold}.}
    \centering
    \resizebox{\textwidth}{!}{%
    \begin{tabular}{ccccccc}
        \hline
         Metric & Model & Gamma ($\gamma$) & Neutron ($n$) & Lambda ($\Lambda$) & K-Short ($K_S^0$) & Sigma+ ($\Sigma^+$) \\
         \hline
         WS & Baseline\_100 & $1.02 \pm 0.02$	& $2.51 \pm 0.03$ & $5.62 \pm 0.29$ & $4.96 \pm 0.14$ & $10.89 \pm 0.68$ \\
         WS & Baseline\_200 & $0.95 \pm 0.03$ & $\mathbf{2.21 \pm 0.03}$ & $\mathbf{5.19 \pm 0.34}$ & $5.07 \pm 0.16$ & $11.09 \pm 0.62$ \\
         WS & FT\_100\_ND & $0.88 \pm 0.01$ & $2.90 \pm 0.04$ & $5.70 \pm 0.19$ & $\mathbf{4.94 \pm 0.15}$ & $\mathbf{9.91 \pm 0.68}$ \\
         WS & FT\_100\_DN & $\mathbf{0.87 \pm 0.01}$ & $2.28 \pm 0.05$ & $5.56 \pm 0.24$ & $5.10 \pm 0.10$ & $10.39 \pm 0.40$ \\
         \hline 
         $\text{MAE}_{cond}$ & Baseline\_100 & $2.01 \pm 0.03$ & $3.94 \pm 0.05$ & $6.96 \pm 0.44$ & $6.03 \pm 0.08$ & $\mathbf{6.37 \pm 0.46}$ \\
         $\text{MAE}_{cond}$ & Baseline\_200 & $1.86 \pm 0.02$ & $\mathbf{3.67 \pm 0.06}$ & $5.95 \pm 0.32$ & $6.01 \pm 0.15$ & $6.51 \pm 0.20$ \\
         $\text{MAE}_{cond}$ & FT\_100\_ND & $\mathbf{1.39 \pm 0.01}$ & $3.85 \pm 0.07$ & $\mathbf{5.78 \pm 0.16}$ & $\mathbf{5.58 \pm 0.13}$ & $7.04 \pm 0.66$ \\
         $\text{MAE}_{cond}$ & FT\_100\_DN & $1.65 \pm 0.01$ & $3.75 \pm 0.12$ & $5.87 \pm 0.42$ & $5.85 \pm 0.15$ & $6.53 \pm 0.29$ \\
         \hline
         $\delta$ & Baseline\_100 & $0.134 \pm 0.003$ & $0.086 \pm 0.001$ & $0.083 \pm 0.004$ & $\mathbf{0.230 \pm 0.006}$ & $0.279 \pm 0.008$ \\
         $\delta$ & Baseline\_200 & $0.124 \pm 0.002$ & $0.081 \pm 0.001$ & $0.079 \pm 0.002$ & $0.238 \pm 0.012$ & $\mathbf{0.263 \pm 0.013}$ \\
         $\delta$ & FT\_100\_ND & $\mathbf{0.098 \pm 0.002}$ & $\mathbf{0.079 \pm 0.002}$ & $\mathbf{0.074 \pm 0.005}$ & $0.237 \pm 0.008$ & $0.275 \pm 0.014$ \\
         $\delta$ & FT\_100\_DN & $0.119 \pm 0.002$ & $0.086 \pm 0.003$ & $0.078 \pm 0.006$ & $0.244 \pm 0.006$ & $0.292 \pm 0.020$ \\
         \hline
         JaccErr & Baseline\_100 & $0.100 \pm 0.001$ & $0.0463 \pm 0.0011$ & $\mathbf{0.079 \pm 0.005}$ & $0.240 \pm 0.005$ & $0.255 \pm 0.016$ \\
         JaccErr & Baseline\_200 & $0.098 \pm 0.001$ & $0.0463 \pm 0.0013$ & $0.081 \pm 0.006$ & $0.247 \pm 0.009$ & $0.248 \pm 0.018$ \\
         JaccErr & FT\_100\_ND & $\mathbf{0.091 \pm 0.001}$ & $\mathbf{0.0455 \pm 0.0016}$ & $0.083 \pm 0.001$ & $\mathbf{0.226 \pm 0.006}$ & $\mathbf{0.228 \pm 0.017}$ \\
         JaccErr & FT\_100\_DN & $0.095 \pm 0.002$ & $0.0470 \pm 0.0015$ & $0.080 \pm 0.003$ & $0.240 \pm 0.007$ & $0.250 \pm 0.007$ \\
         \hline
    \end{tabular}}
    \label{tab:results-orgps}
\end{table}

To assess whether the differences between models are statistically significant, we performed bootstrap resampling of the unique-input level for the metrics $\text{MAE}_{cond}$, $\delta$, and JaccErr. We focus on FT\_100\_ND because it usually performs better than FT\_100\_DN, and on the Baseline\_200. We exclude $\Sigma+$ due to small number of samples available. First, we computed each metric at the unique-input level and averaged the results over five runs. Then, for each of the $10{,}000$ bootstrap samples, we resampled unique inputs, preserving all associated responses and weights, and computed the differences between the fine-tuned and baseline models. This yielded p-values, 95\% confidence intervals, and the fraction of bootstrap samples favouring the fine-tuned model. The results are shown in Table~\ref{tab:results-orgps-bootstrap}. For the $\gamma$ particles, all fine-tuned results are significantly better than the baselines. This suggests that, when enough particle-specific samples are available, fine-tuning is the most efficient way to obtain a surrogate model. For the other cases, the conclusions are less clear: some results are statistically significant in favour of the fine-tuned models, while none favour the baseline. Whenever the fine-tuned models outperform the baseline in Table~\ref{tab:results-orgps}, the majority of bootstrap samples follow the same trend. However, this analysis should be interpreted cautiously. In some cases, only a small number of samples is available per unique input. In addition, resampling the observed unique inputs changes the effective input distribution, which may not correspond to a physically plausible perturbation.

\begin{table}[ht]
    \caption{Comparison between FT\_100\_ND and the Baseline\_200 models for different particles based on bootstrap samples. Results with $\text{p-value} < 0.05$ are highlighted in \colorbox{ForestGreen}{dark green}, and those close to $\text{p-value}=0.05$ -- in \colorbox{LimeGreen}{green}.}
    \centering
    \begin{tabular}{cccccc}
    \hline
         Metric & Result type & Gamma ($\gamma$) & Neutron ($n$) & Lambda ($\Lambda$) & K-Short ($K_S^0$) \\
         \hline
         $\text{MAE}_{cond}$ & CI & \cellcolor{ForestGreen}-0.603 -0.347 & -0.093 0.469 & -1.334 0.805 & \cellcolor{ForestGreen}-0.824 -0.025 \\
         $\text{MAE}_{cond}$ & P, P\_val & \cellcolor{ForestGreen}1.000, 0.000 & 0.104, 0.208 & 0.620, 0.761 & \cellcolor{ForestGreen}0.980, 0.041 \\
         \hline
         $\delta$ &  CI & \cellcolor{ForestGreen}-0.030 -0.021 & -0.005 0.002 & \cellcolor{LimeGreen}-0.011 0.000 & -0.014 0.011 \\
         $\delta$ & P, P\_val & \cellcolor{ForestGreen}1.000, 0.000 & 0.853, 0.295 & \cellcolor{LimeGreen}0.973, 0.054 & 0.589, 0.821 \\
         \hline
         JaccErr & CI & \cellcolor{ForestGreen}-0.010 -0.005 & -0.003 0.002 & -0.006 0.009 & \cellcolor{ForestGreen}-0.033 -0.010 \\
         JaccErr & P, P\_val & \cellcolor{ForestGreen}1.000, 0.000 & 0.712, 0.576 & 0.342, 0.685 & \cellcolor{ForestGreen}1.000, 0.000 \\
         \hline 
    \end{tabular}
    \label{tab:results-orgps-bootstrap}
\end{table}
\vspace{-15pt}

\subsubsection{Two-stage modelling: the full pipeline analysis}

Here, we analyse the performance of the full two-stage pipeline. The results for all models and metrics are shown in Table~\ref{tab:results-wholepipeline}. For each particle and setup, we report scores obtained using the best-performing noise-removal method with respect to the Wasserstein score. Because training models from scratch for each particle performs significantly worse than other approaches, it is excluded from the analysis.


\vspace{-12pt}
\begin{table}[ht]
    \caption{Comparison of model performances for different particles and metrics. \textit{Baseline} denotes models trained on the full dataset and \textit{FT\_100} denotes fine-tuned Baseline\_100 models. Superscripts indicate the noise removal method used. The best results are in \textbf{bold}.}
    \centering
    \resizebox{\textwidth}{!}{%
    \begin{tabular}{ccccccc}
        \hline
         Metric & Model & Gamma ($\gamma$) & Neutron ($n$) & Lambda ($\Lambda$) & K-Short ($K_S^0$) & Sigma+ ($\Sigma^+$) \\
         \hline
         WS & Baseline\_100 & $1.80 \pm 0.02^3$ & $\mathbf{3.47 \pm 0.07}^1$ & $16.23 \pm 0.22^1$ & $\mathbf{10.51 \pm 0.06}^2$ & $33.94 \pm 0.35^1$ \\
         WS & Baseline\_200 & $1.75 \pm 0.02^3$  & $3.85 \pm 0.05^1$ & $16.83 \pm 0.16^1$ & $11.03 \pm 0.14^2$ & $32.38 \pm 0.46^1$ \\
         WS & FT\_100\_ND & $\mathbf{1.58 \pm 0.01}^3$ & $4.53 \pm 0.05^3$ & $\mathbf{15.79 \pm 0.21^1}$ & $10.94 \pm 0.23^2$ & $\mathbf{31.52 \pm 0.64}^1$ \\
         WS & FT\_100\_DN & $1.70 \pm 0.02^3$ & $3.56 \pm 0.07^1$ & $16.00 \pm 0.26^1$ & $10.56 \pm 0.20^2$ & $32.62 \pm 0.41^1$ \\
         \hline
         $\text{MAE}_{cond}$ & Baseline\_100 & $4.89 \pm 0.02^3$ & $7.04 \pm 0.03^1$ & $10.63 \pm 0.42^1$ & $9.77 \pm 0.17^2$ & $17.23 \pm 0.26^1$ \\
         $\text{MAE}_{cond}$ & Baseline\_200 & $4.78 \pm 0.02^3$ & $6.95 \pm 0.08^1$ & $9.70 \pm 0.18^1$ & $\mathbf{9.59 \pm 0.03}^2$ & $\mathbf{16.63 \pm 0.22}^1$ \\
         $\text{MAE}_{cond}$ & FT\_100\_ND & $\mathbf{4.53 \pm 0.01}^3$ & $7.29 \pm 0.03^3$ & $9.73 \pm 0.19^1$ & $10.11 \pm 0.13^2$ & $17.11 \pm 0.33^1$ \\
         $\text{MAE}_{cond}$ & FT\_100\_DN & $4.65 \pm 0.02^3$ & $\mathbf{6.78 \pm 0.08}^1$ & $\mathbf{9.43 \pm 0.08}^1$ & $9.88 \pm 0.19^2$ & $17.63 \pm 0.14^1$  \\
         \hline
         $\delta$ & Baseline\_100 & $0.146 \pm 0.001^3$ & $\mathbf{0.145 \pm 0.004}^1$ & $\mathbf{0.148 \pm 0.005}^1$ & $0.306 \pm 0.006^2$ & $0.337 \pm 0.008^1$ \\
         $\delta$ & Baseline\_200 & $0.137 \pm 0.002^3$ & $0.152 \pm 0.002^1$ & $0.154 \pm 0.006^1$ & $0.296 \pm 0.006^2$ & $0.334 \pm 0.013^1$ \\
         $\delta$ & FT\_100\_ND & $\mathbf{0.120 \pm 0.001}^3$ & $0.184 \pm 0.001^3$ & $0.152 \pm 0.002^1$ & $\mathbf{0.286 \pm 0.008}^2$ & $\mathbf{0.328 \pm 0.010}^1$ \\
         $\delta$ & FT\_100\_DN & $0.133 \pm 0.002^3$ & $0.154 \pm 0.002^1$ & $0.150 \pm 0.004^1$ & $0.289 \pm 0.008^2$ & $0.333 \pm 0.013^1$ \\
         \hline
         JaccErr & Baseline\_100 & $0.113 \pm 0.001^3$ & $0.0570 \pm 0.0016^1$ & $0.100 \pm 0.003^1$ & $0.245 \pm 0.006^2$ & $0.313 \pm 0.015^1$ \\
         JaccErr & Baseline\_200 & $0.111 \pm 0.001^3$ & $\mathbf{0.0543 \pm 0.0009}^1$ & $0.093 \pm 0.004^1$ & $0.247 \pm 0.007^2$ & $0.305 \pm 0.007^1$  \\
         JaccErr & FT\_100\_ND & $\mathbf{0.105 \pm 0.002}^3$ & $0.0559 \pm 0.0020^3$ & $\mathbf{0.090 \pm 0.006}^1$ & $\mathbf{0.236 \pm 0.006}^2$ & $\mathbf{0.276 \pm 0.010}^1$ \\
         JaccErr & FT\_100\_DN & $0.109 \pm 0.001^3$ & $0.0544 \pm 0.0016^1$ & $0.093 \pm 0.007^1$ & $0.245 \pm 0.007^2$ & $0.299 \pm 0.022^1$ \\
         \hline
    \end{tabular}}
    \label{tab:results-wholepipeline}
\end{table}

Fine-tuning outperforms all other approaches for $\gamma$, $\Lambda$, and $\Sigma^+$ in terms of Wasserstein score. For $K_S^0$ and $n$, fine-tuning results are only slightly worse than the baseline.



The best noise-removal method is consistent between baseline and fine-tuned models for each particle. The only exception is $n$, where floor-based methods with and without recalculation perform best in different settings. This suggests that the model behaviour is consistent and that a once-chosen noise-removal method for a given particle should also work well for similar models.


Compared to Table~\ref{tab:results-orgps}, the full pipeline shows worse metric values, especially for $\Lambda$, $K_S^0$, and $\Sigma^+$. This indicates that the auxiliary photon-count predictor is the main bottleneck, as it fails to learn the input-to-photon mapping well for these particles.



The other metrics also generally favour fine-tuning. Interestingly, FT\_100\_DN sometimes performs best, unlike in the previous NF-only analysis. The metric that remains most consistent between Tables~\ref{tab:results-orgps} and~\ref{tab:results-wholepipeline} is Jaccard index error; the only significant change in the best-performing model is for the $\Lambda$ case, which is not surprising given the much worse Wasserstein score observed there.


The most stable results are again obtained for $\gamma$, which is the least diverse particle type and also the most abundant in the data set. Here, FT\_100\_ND outperforms the rest. For $\Sigma^+$, FT\_100\_ND is the best in 3 of the 4 metrics and second best in the remaining one, which also makes it stand out. For the remaining cases, the ranking is less straightforward. We therefore rank the results within each metric and combine them into a single averaged value. For $n$, Baseline\_200 and FT\_100\_DN perform similarly, although Baseline\_200 is only slightly better in Jaccard index error, which suggests that the FT\_100\_DN can be considered to be better overall. For $\Lambda$, FT\_100\_ND and FT\_100\_DN also have the same mean ranking, and the slight advantage of FT\_100\_DN becomes visible only when Table~\ref{tab:results-orgps} is taken into account. For $K_S^0$, Baseline\_100 and FT\_100\_ND have the same ranking, but the fine-tuned model is preferable when Table~\ref{tab:results-orgps} is considered as well.


Across the four better-sampled particle types ($\gamma$, $n$, $\Lambda$, $K_S^0$) and the four metrics, the fine-tuned FT\_100\_ND model is best overall in most comparisons. The difference between results presented in Tables~\ref{tab:results-orgps} and~\ref{tab:results-wholepipeline} suggests that the main bottleneck in this study is the prediction of the expected photon number, so further work should focus on that stage.

We also want to highlight that fine-tuning is more efficient than training the baseline longer on the full dataset. Thus, in cases where the fine-tuning improves the Baseline\_200 or where it behaves similarly, it makes more sense to follow the fine-tuning approach, which saves time and reduces energy consumption.


\subsection{Ensemble model evaluation}

\begin{figure}[ht]
\centering
\includegraphics[width=\linewidth]{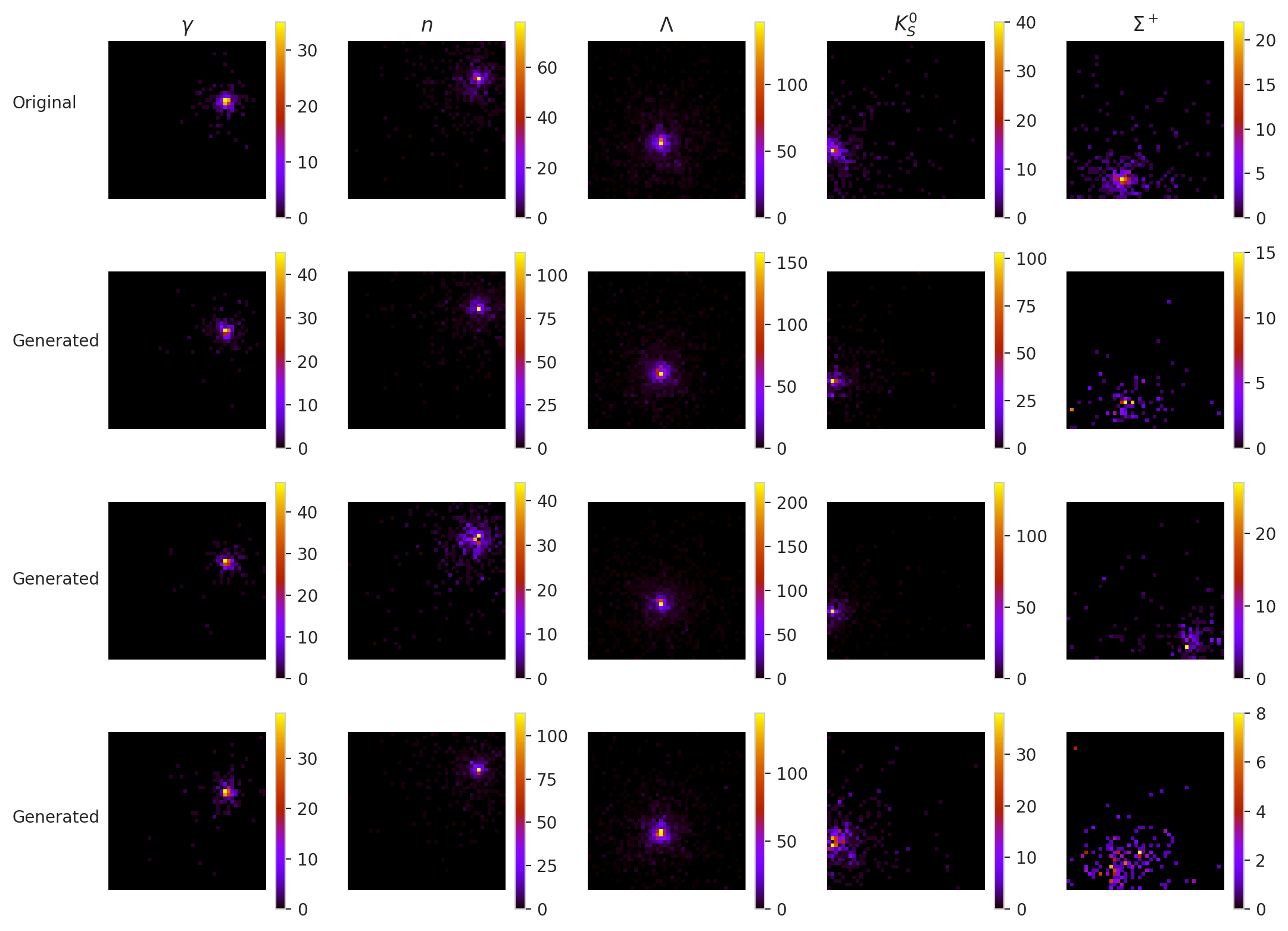}
\caption{Sample results from the ensemble model. Three samples per input vector illustrate run-to-run variability.} 
\label{fig:flow-result-ft}
\end{figure}

We construct the final ensemble by selecting the best-performing model for each particle: FT\_100\_ND for $\gamma$, $K_S^0$, and $\Sigma^+$; FT\_100\_DN for $n$ and $\Lambda$; and Baseline\_100 for the remaining particles (six models total). Table~\ref{tab:final-comparison} compares the ensemble against the baselines across all metrics. The ensemble outperforms both baselines, validating the multi-model approach. Sample ensemble outputs are shown in Fig.~\ref{fig:flow-result-ft}.


\vspace{-15pt}
\begin{table}[ht]
    \caption{Final comparison of the ensemble model against baselines, computed over the full test dataset.}
    \centering
    \begin{tabular}{cccccc}
        \hline
         & WS & $\text{MAE}_{cond}$ & $\delta$ & $\text{JaccErr}$  \\
         \hline
         Baseline\_100 & $1.70 \pm 0.02$ & $6.57 \pm 0.04$ & $0.176 \pm 0.001$ & $0.127 \pm 0.000$\\
         Baseline\_200 & $1.76 \pm 0.02$ & $6.43 \pm 0.03$ & $0.172 \pm 0.001$ & $0.125 \pm 0.001$ \\
         Ensemble & $\mathbf{1.61 \pm 0.02}$ & $\mathbf{6.29 \pm 0.02}$ & $\mathbf{0.163 \pm 0.001}$ & $\mathbf{0.121 \pm 0.001}$ \\
         \hline
    \end{tabular}
    \label{tab:final-comparison}
\end{table}

\vspace{-10pt}



\subsection{Time performance}

MAF-based NF architectures are fast in evaluating the likelihood, which allows for reasonably short training, and slow during sampling. Our models require around \SI{160.0}{\ms} for response generation, compared to e.g. \SI{0.023}{\ms} needed for a GAN (NVIDIA A100 40 GB GPU, batch size 256)~\cite{zdcfastsim}. However, it is possible to train an Inverse Autoregressive Flow (IAF) student, which mimics the behaviour of the previously trained MAF teacher. This way, we can expect to obtain around 420 times faster models (\SI{0.38}{\ms} per sample), as in~\cite{Majerz:2025ykn}.



\subsection{Understanding the model}

Good performance alone is insufficient; models should also exhibit physically plausible reasoning. We use SHAP analysis to verify the ensemble's feature dependencies for three quantities: total photon count, and shower centre $x$ and $y$ coordinates. For photon count, we explain the BNN predictions directly. For coordinates, we train surrogate BNNs on NF-generated data through knowledge distillation~\cite{hinton2015distillingknowledgeneuralnetwork}, as a direct NF explanation is too slow.



The analysis indicates that the photon count depends primarily on \textit{Energy} and momentum in the $z$ direction $P_z$, with higher values producing more photons, as expected physically. Because we removed the sign from $P_z$ (indicating ZDC side, negligible here), its contribution is similar to \textit{Energy} due to the energy-momentum relation $E^2 = P_x^2 + P_y^2 + P_z^2 + mass^2$ (in the dataset, the $P_z$ value is orders of magnitude bigger than $P_x$, $P_y$ and $mass$, thus dominating the $E$ value). Shower $x$ and $y$ coordinates depend mainly on $P_x$, $P_y$, and \textit{Energy}, as expected from particle trajectory physics. The influence of \textit{charge} reflects magnetic field deflection near the detector.

\section{Conclusions}

We applied Normalizing Flows (NFs) to simulate ZDC neutron calorimeter responses, demonstrating substantial improvements through transfer learning, fine-tuning, and ensembling. Our ensemble generator achieves a Wasserstein distance of $1.61 \pm 0.02$, outperforming all baselines across all metrics while maintaining physically plausible feature dependencies verified via SHAP analysis. Importantly, fine-tuning saves time and reduces energy consumption compared to a longer training on the full dataset. 


Standard ZDC evaluation metrics, such as Wasserstein distance and per-pixel MAE, have known limitations. We therefore introduce conditional metrics: weighted channel MAE, dispersion ratio, and Jaccard co-activation error, which better capture input-output physical consistency across diverse particle types and response variability.




Key methodological advances include:
\begin{enumerate}
    \item Two fine-tuning schemes using gradual unfreezing, with one of them showing the strongest overall performance.
    \item Four post-processing methods for noise removal, where particle-specific optimization yields further gains.
    \item An ensemble combining specialised models that validates the multi-model approach.
\end{enumerate}



The analysis reveals that photon-count prediction remains the primary bottleneck, suggesting clear directions for future improvement. While current inference is slower than some other generative alternatives, IAF distillation offers a path to a substantial speedup. We also highlight the possibility of utilising different approaches to ensemble modelling, e.g., with a meta-model designed to assign a simulation task to a particular generative surrogate.

This work provides a generalizable procedure for constructing a data-driven surrogate model of a complex numerical simulation, specifically aimed at predicting detector responses to particle parameters recorded at the LHC. The combination of normalizing flows, conditional fine-tuning, and physics-motivated evaluation establishes a strong foundation for production-ready fast simulation.




\begin{credits}
\subsubsection{\ackname} We would like to thank Sandro Wenzel, PhD from CERN for his support in this work. This work is co-financed and in part supported by the Ministry of Science and Higher Education (Agreements No.~2022/WK/01 and 2023/WK/07) by the program entitled ``PMW'' and by the Ministry funds assigned to AGH University of Krakow. We gratefully acknowledge Polish high-performance computing infrastructure PLGrid (HPC Center: ACK Cyfronet AGH) for providing computer facilities and support within computational grants no.~PLG/2023/016410, PLG/2024/017264 and PLG/2025/018322.

\subsubsection{\discintname}
The authors have no competing interests to declare that are relevant to the content of this article.
\end{credits}
%
%
%
\bibliographystyle{splncs04}
\bibliography{bibliography}

\end{document}